\documentclass[twocolumn,showpacs,preprintnumbers,amsmath,amssymb]{revtex4}
\usepackage{graphicx}
\usepackage{dcolumn}
\usepackage{bm}
\usepackage{comment}
\usepackage{rotating}
\usepackage{longtable}
\usepackage{float}
\usepackage{eucal}
\usepackage{csquotes}


\begin{document}

\title{\centering {Extremely asymmetric shears band in $^{143}$Sm}}

\author{S. Rajbanshi$^{1}$}
\email{subhphy@gmail.com}
\author{R. Raut$^{2}$}

\author{H. Pai$^{3}$}
\email{h.pai@saha.ac.in}
\author{Sajad Ali$^{3}$}
\author{A. Goswami$^{3}$}

\author{S. Bhattacharyya$^{4}$}
\author{G. Mukherjee$^{4}$}
\author{R. K. Bhowmik$^{5}$}
\author{S. Muralithar$^{5}$}
\author{R. P. Singh$^{5}$}
\author{G. Gangopadhyay$^{6}$}
\author{M. Kumar Raju$^{7}$}
\author{P. Singh$^{8}$}

\affiliation{$^{1}$Department of Physics, Presidency University, Kolkata 700073, India}
\affiliation{$^{2}$UGC-DAE Consortium for Scientific Research, Kolkata Centre, Kolkata 700098, India}
\affiliation{$^{3}$Saha Institute of Nuclear Physics, Kolkata 700064, India}
\affiliation{$^{4}$Variable Energy Cyclotron Center, Kolkata 700064, India}
\affiliation{$^{5}$Inter University Accelerator Centre, Aruna Asaf Ali Marg, New Delhi 110067, India}
\affiliation{$^{6}$Department of Physics, University of Calcutta, Kolkata 700009, India}
\affiliation{$^{7}$Research Center for Nuclear Physics, Osaka University, Japan}
\affiliation{$^{8}$IRFU, CEA, Université Paris-Saclay, F-91191 Gif-sur-Yvette, France}

\date{\today}

\begin{abstract}

A dipole sequence has been observed and investigated in the $^{143}$Sm nucleus populated through the heavy-ion induced fusion-evaporation reaction and studied using the Indian National Gamma Array (INGA) as the detection system. The sequence has been established as a Magnetic Rotation (MR) band primarily from lifetime measurements of the band members using the Doppler Shift Attenuation Method (DSAM). A configuration based on nine quasiparticles, with highly asymmetric angular momentum blades, has been assigned to the shears band in the light of the theoretical calculations within the framework of Shears mechanism with the Principal Axis Cranking (SPAC) model. This is hitherto the maximum number of quasiparticles along with the highest asymmetricity associated with a MR band. Further, as it has followed from the SPAC calculations, the contribution of the core rotation to the angular momentum of this shears band is substantial and greater than in any other similar sequence, at least in the neighbouring nuclei. This band can thus be perceived as a unique phenomenon of shears mechanism in operation at the limits of quasiparticle excitations, as manifested in MR band-like phenomena, evolving into collectivity.

\end{abstract}


\maketitle

Recent experimental and theoretical studies on the weakly deformed nuclei, with a very few particles and holes outside the core, unambiguously establish \cite{ajsim, hubel, rmcla1,amita} the \enquote{shears mechanism} as a general phenomenon of generating angular momentum in them. The shears mechanism in these nuclei, manifested in form of the magnetic rotational (MR) bands, has been observed in their level structures. These bands are characterized by strong intraband M1 transitions and weak/unobserved cross-over E2 transitions, the latter being commensurate with the small deformations of these nuclei. The band-head corresponds to the perpendicular alignment of the angular momentum vectors generated by the particle and the holes constituting the band configuration. The repulsive interaction between the particle and holes favours the perpendicular coupling, for minimum energy at the band-head. Excited states with higher angular momenta along the MR band are generated by the gradual alignment of the angular momentum vectors that may eventually completely align to produce the maximum spin accompanied by the termination of the MR band \cite{frauen1}.

The mid-shell nuclei, on the other hand, have considerable number of valence nucleons outside the core that is consequently deformed owing to the polarizing effect of the former. The deformation breaks the rotational symmetry and leads to the observation of the so-called rotational bands in the level structure of the corresponding nuclei. This may be perceived as a transition in the characteristic excitation pattern of the nuclei with increasing number of valence nucleons outside the core. The weakly deformed systems with a few valence particles exhibiting MR-bands, or similar single particle features, evolves into well deformed  nuclei with increased valence population and exhibit collectivity in the form of rotational sequences. In such a transitional picture, observation of shears sequences based on large number of quasiparticles may actually represent an interesting step in the aforementioned structural progression, before the onset of collective rotational excitations. The systematic investigation of such an evolutionary scenario is warranted.    

MR bands have been observed in several mass regions, $A$ $\approx$ 80, 100, 140, 190 \cite{ajsim, hubel, rmcla1,amita}, across the nuclear chart. The highest number of quasiparticles in the shears configuration has been established for sequences observed in $^{108}$Cd \cite{108cd}, $^{198}$Pb \cite{199pb}, and $^{198}$Bi \cite{198bi} of which that in the $^{198}$Bi is the most asymmetric one; the asymmetry being quantified by the difference in angular momenta of the two shears blades, $j_{p}$ and $j_{n}$. There have been numerous observation of asymmetric shears configurations in the literature albeit the limits of such asymmetry that would (still) favor the shears mechanism for generation of angular momentum remains to be explored. Such an objective is an impetus to the present endeavor that consists of a quest for asymmetric shears configurations in $^{143}$Sm [$Z$ = 62, $N$ = 81]. The nucleus is characterized by enhanced probability of particle excitations in the proton sector while a hindered one in case of the neutrons, owing to the $N$ = 82 closure, that may actually result in the sought asymmetry.

In the light of the aforementioned prospects, the $^{143}$Sm ($Z = 62, N = 81$) nucleus has been investigated in the present work. Recently, two dipole bands, DB I and DB II, have been reported in the level structure of the nucleus above an excitation energy of $E_{x}$ = 8614-keV \cite{rajban4, rajban6, rraut}. Recent investigation reveals another dipole band almost degenerate in energy with the DB II without measurements of polarization and transition rates leading the interpretation only tentative \cite{rajban7}. The current work re-investigates this third dipole band in the nucleus which has been generated from the extremely asymmetric shears structure on the \enquote{phase boundary} of the two phases representing the quasiparticle excitation and deformed core rotation. The proposition has been validated through level lifetime measurements of the band members and extracting $B(M1)$, and $B(E2)$ values and their evolution, along the band, therefrom. The experimental findings have been well reproduced within the framework of a modified shears mechanism with principal axis cranking (SPAC) model to further establish/uphold the interpretation.

The dipole structures above 8614-keV 43/2$^{-}$ state in $^{143}$Sm were populated using heavy-ion fusion evaporation reaction $^{124}$Sn ($^{24}$Mg, 5n) at the beam ($^{24}$Mg) energy of 107-MeV provided by the 15UD pelletron facility at Inter University Accelerator Center (IUAC), New Delhi. Target was 0.8 mg/cm$^2$ thick $^{124}$Sn [99.9\% enriched] evaporated on a 13 mg/cm$^2$ gold backing. The recoils were produced with $\beta$ $\approx$ 1.6\%. The deexciting $\gamma$-ray transitions were detected using the Indian National Gamma Array (INGA) \cite{smurali} which consisted of eighteen Compton suppressed clover detectors arranged in six different angles [90$^{\circ}$(6), 123$^{\circ}$(4), 148$^{\circ}$(4) and 57$^{\circ}$(4)] with respect to the beam axis (the number in the parenthesis is the detector numbers at the respective angles). About 8$\times$10$^{8}$ two and higher fold $\gamma$-$\gamma$ coincidence events were collected in the list mode format. The data was sorted into different symmetric and angle dependent $E_{\gamma}$ - $E_{\gamma}$ matrices with the help of the INGASORT and analyzed using the INGASORT and the RADWARE packages \cite{ingasort, radford1, radford2}.

\begin{figure}[t]
\centering
\setlength{\unitlength}{0.05\textwidth}
\begin{picture}(10,12.9)
\put(0.2,14.6){\includegraphics[width=0.35\textwidth, angle = -90]{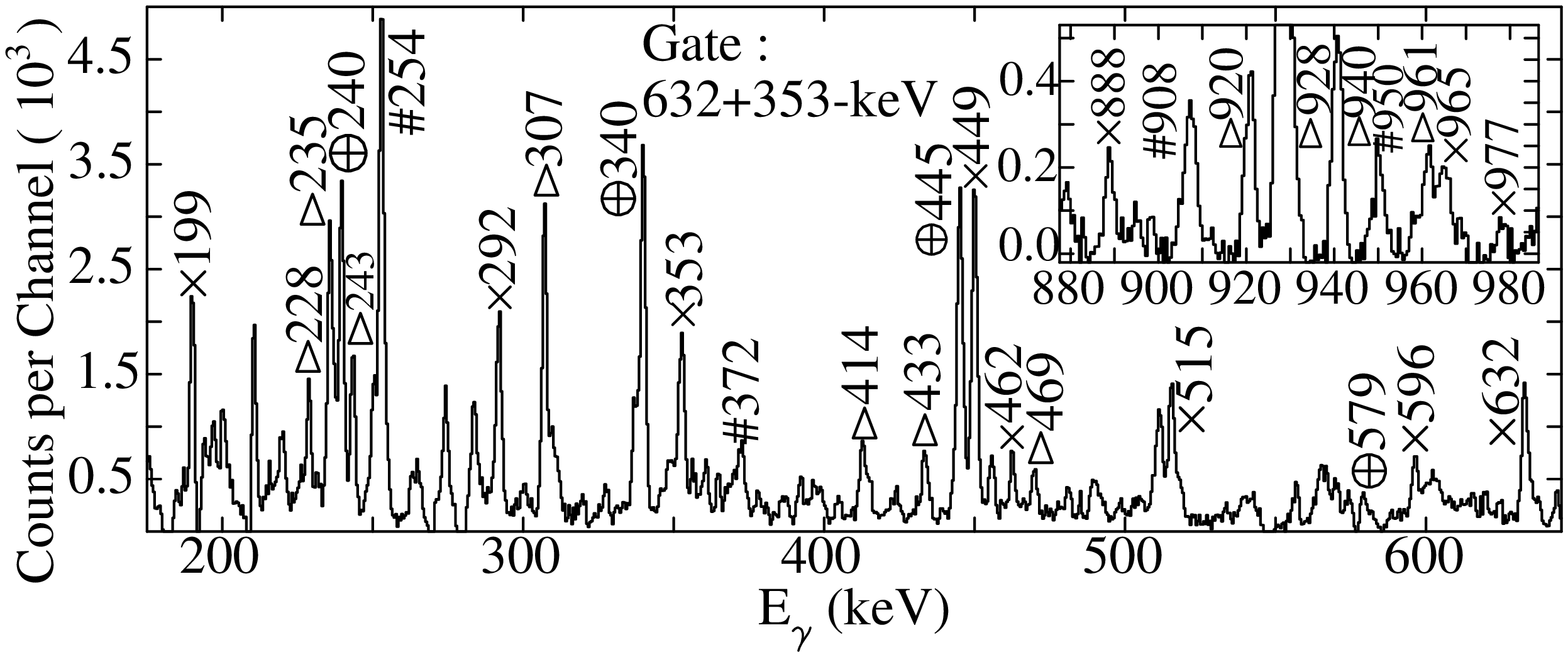}}
\put(-1.7,9.5){\includegraphics[width=0.51\textwidth, angle = -90]{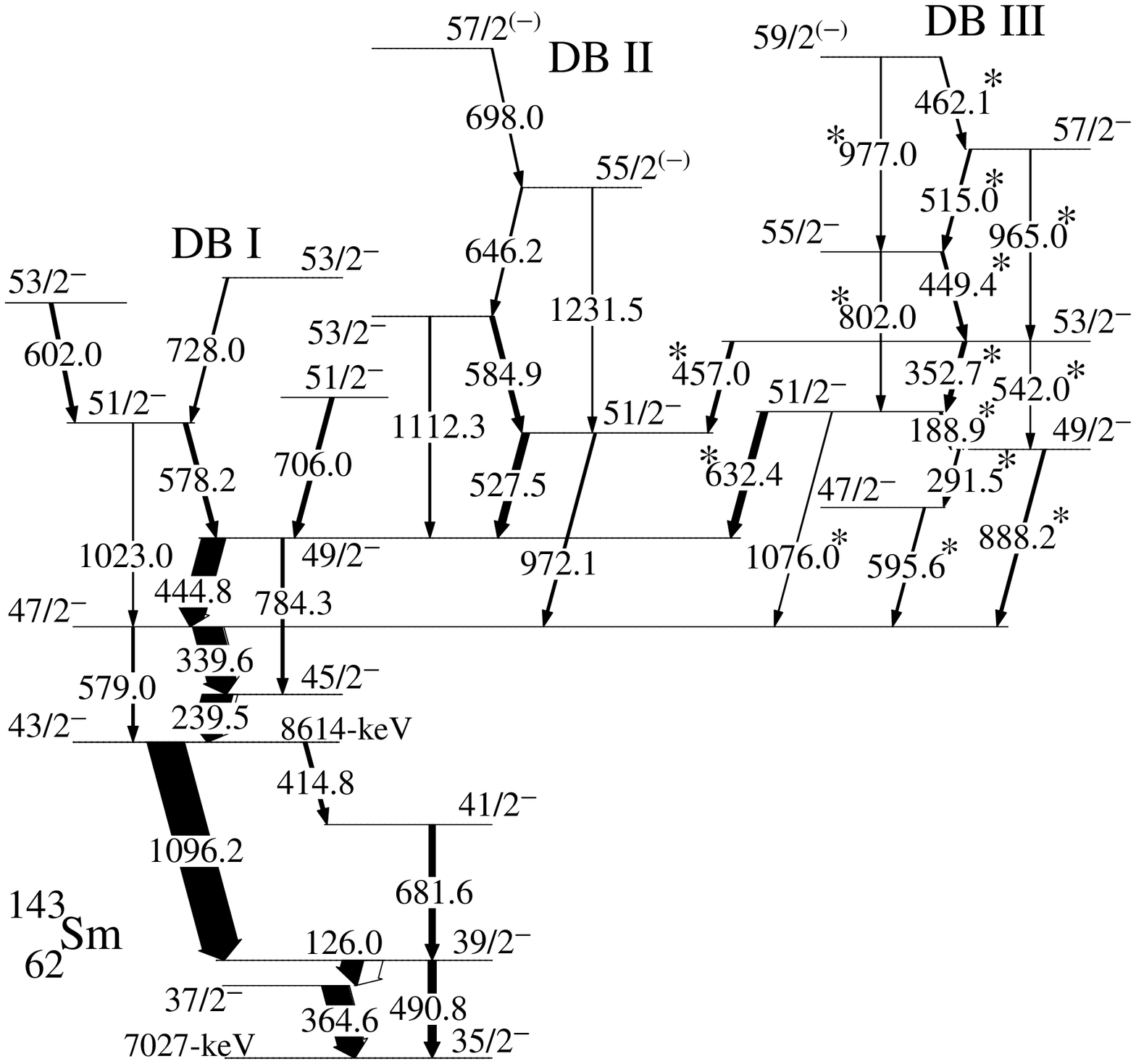}}
\put(5.1,-0.0){\includegraphics[width=0.22\textwidth, angle = 0]{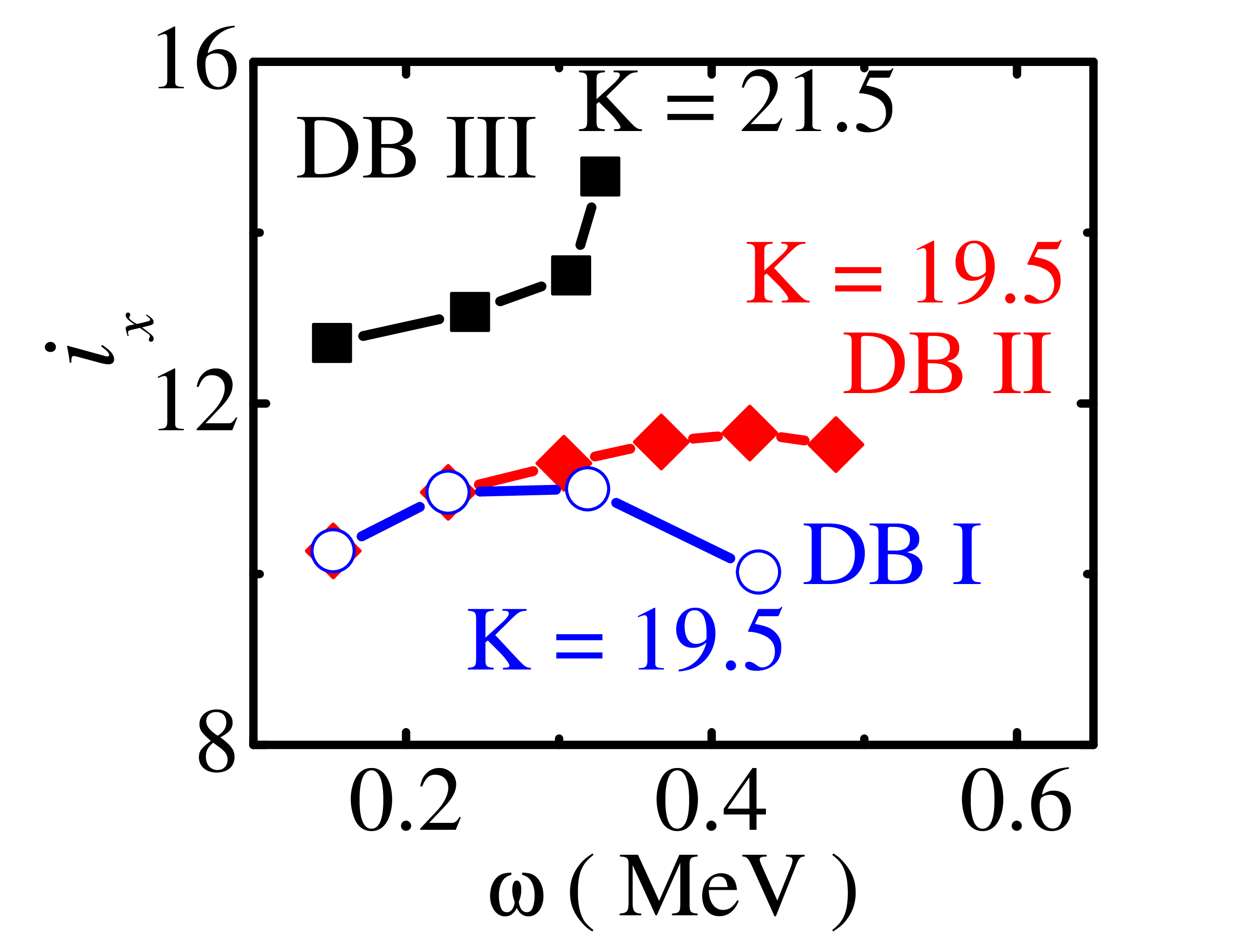}}
\put(0.2,-0.2){\textbf{(a)}}
\put(5.5,0.2){\textbf{(c)}}
\put(0.5,9.20){\textbf{(b)}}

\end{picture}
\caption{\label{levsch} (Color online.) (a) Proposed negative parity level structure above the 7027-keV 35/2$^{-}$ state in $^{143}$Sm. The newly observed $\gamma$ -rays are marked by an asterisk. (b) Summed gated spectrum of the 352.7 and 632.4-keV transitions shows $\gamma$ rays marked by \enquote{$\times$} (DB III), \enquote{$\oplus$} (DB I) and \enquote{$\Delta$} (below 8614-keV) in $^{143}$Sm. The contamination peaks from the $^{144}$Sm, populated in the same reaction, has been marked with \enquote{$\#$}.  Aligned angular momentum ($i_{x}$) of the bands is shown in (c). Harris parameters used in the calculation of $i_{x}$ are $\it{J_{0}}$ = 12$\hbar^2$ MeV$^{-1}$ and $\it{J_{1}}$ = 25$\hbar^4$ MeV$^{-3}$ \cite{dmcull}. The band-head spin (K) is determined from the fitting of excited energy ($E(I)$) against spin ($I$).}
\end{figure}

The multipolarities and the electromagnetic characters of the observed $\gamma$ -ray transitions, for assigning the spin-parity of the levels, were determined from the measurements of the of ratio for Directional Correlation from Oriented state ($R_{DCO}$) \cite{kramer, kabadi}, linear polarization asymmetry ($\Delta$) \cite{staro2, droste2, deng2, jones2}  and the mixing ratio ($\delta$) \cite{tyamaza, edmawsu, macias}. The experimental details and data analysis procedures have been described in Ref. \cite{rajban, rajban3}.

\begin{table*}
\centering
\caption{\label{lifetime} The DCO ratio ($R_{DCO}$), polarization asymmetry ($\Delta$), mixing ratio ($\delta$), level lifetimes ($\tau$) and side feeding lifetimes ($\tau_{sf}$) of the states and the corresponding $B(M1)$ and $B(E2)$ transitions rates for the $\gamma$ transitions of the dipole band DB III in $^{143}$Sm. The uncertainties are rounded off to the nearest value up to two decimal places. Correction in the transition strength due to the internal conversion has been incorporated for the $\gamma$ transitions having energy ($E_{\gamma}$) less than 400.0-keV.}

\begin{tabular}{cccccccccccccc}

\hline\hline

$J_{i}$      &$J_{f}$    & $E_\gamma$ & Assign. & $I_{\gamma}$$^{a}$ & $R_{DCO}$$^{b}$ & $\Delta$ & $\delta$  & $\tau$   & $\tau_{sf}$ &  $B(M1)$            &     $B(E2)$    \\

[ $\hbar$ ]  & [ $\hbar$ ]& [ keV ]    &  &                  &                 &          &           & [ ps ]   & [ ps ] & [ $\mu_{N}^{2}$ ]  &  [ $e^2b^2$ ]  \\

\hline\hline

47/2$^{-}$ & 47/2$^{-}$ & 595.6(4)  & $\Delta$I = 0, M1 & 6.7(12)    & 1.68(19)   & +0.10(9)  &        &                       &                          &      & \\

49/2$^{-}$   & 47/2$^{-}$ & 888.2(3) & M1/E2  & 9.5(7)    & 1.04(8)   & -0.20(5)  & -0.03(5) &                       &                          &      &                      \\
             & 47/2$^{-}$ & 291.5(3) & M1/E2  & 6.4(8)    & 1.18(13)   & -0.19(9)        &        &                       &                          &      & \\

51/2$^{-}$  & 49/2$^{-}$ & 188.9(3) & M1/E2  &  6.5(4)    &  0.93(10) &           & -0.10(7) &  1.17$^{+0.33}_{-0.26}$ & 0.16(4) & 1.17$^{+0.39}_{-0.34}$ &      \\

            & 49/2$^{-}$ & 632.4(4) & M1/E2 & 21.6(10)   & 1.07(7)   & -0.19(8)  & +0.01(4) &       & & 0.14$^{+0.05}_{-0.04}$&          \\

            & 47/2$^{-}$ & 1076.0(5) & E2 & 2.5(4)     & 1.72(18)  &           &          &       &          &  & 0.01(1)\\

53/2$^{-}$  & 51/2$^{-}$ & 352.7(3) & M1/E2  & 13.5(7)   &  1.10(10) & -0.22(12) & +0.02(3)  & 0.86$^{+0.13}_{-0.12}$ & 0.15(3) & 0.74$^{+0.11}_{-0.10}$ &                    \\

            & 49/2$^{-}$ & 542.0(4) & E2 & 1.7(2)    &  1.68(18) &           &          &                         &                        & & 0.13$^{+0.02}_{-0.02}$ \\

            & 51/2$^{-}$ & 457.0(3) & M1/E2  &  11.1(8)     & 0.74(6) &-0.33(15) & -0.28(7) &                  &       & 0.27$^{+0.06}_{-0.06}$                  &\\

55/2$^{-}$  & 53/2$^{-}$ & 449.4(3) & M1/E2 & 9.8(5)    &  0.74(7)  & -0.18(9)  & -0.28(8)  & 0.55$^{+0.08}_{-0.08}$ & 0.13(3) & 0.80$^{+0.12}_{-0.12}$  &                   \\

            & 51/2$^{-}$ & 802.0(5) & E2  & 3.1(3)    &  1.79(18) &           &           &                        &                         & & 0.11$^{+0.02}_{-0.02}$ \\

57/2$^{-}$  & 55/2$^{-}$ & 515.0(3) & M1/E2  & 7.7(4)    &  0.67(7)  & -0.18(9)  & -0.37(12) & 0.43$^{+0.07}_{-0.07}$ & 0.11(2) & 0.58$^{+0.09}_{-0.09}$ &                       \\

            & 53/2$^{-}$ & 965.0(4) & E2  & 3.6(2)    &  1.65(16) &            &           &                        &                        & & 0.07$^{+0.01}_{-0.01}$ \\

59/2$^{(-)}$  & 57/2$^{-}$ & 462.1(3) & M1/E2  & 5.8(3)    &  0.90(8)  &            & -0.13(7)  & 0.62$\downarrow$   & 0.09(2) & 0.63$\uparrow$          &                        \\

            & 55/2$^{-}$ & 977.0(5) & E2  & 2.6(4)    &        1.59(16) &           &           &                         &                      &  & 0.05$\uparrow$      \\

\hline\hline

\multicolumn{11}{l}{$^{a}$Intensities are normalized with the intensity of the 239.5-keV (45/2$^{-}$ $\rightarrow$ 43/2$^{-}$) $\gamma$-transition as 100.}\\

\multicolumn{11}{l}{$^{b}$$R_{DCO}$ [$I_{\gamma}$(148$^{\circ}$)/$I_{\gamma}$(90$^{\circ}$)] values are obtained form the 239.5 [$\delta$ = -0.06(3)] and 339.6 [$\delta$ = -0.11(3)] keV dipole gates.}\\

\end{tabular}

\end{table*}

The present work has been confirmed all the previously observed states and the decay out $\gamma$ transitions of the dipole structure as depicted in Fig. \ref{levsch} (a). A representative sum gated spectrum exhibits some of the $\gamma$-ray transitions, reported by the earlier works along with the newly observed in the present investigation, belonging to the level structure of $^{143}$Sm has been illustrated in Fig. \ref{levsch} (b). As already reported in Ref. \cite{rajban6}, the present measurement, carried out with increased number of detectors, reveals that the 602.0 and 706.0 keV transitions are in parallel with the 728.0 and 578.2-keV transitions of DB I, respectively, thereby placed them above the 51/2$^{-}$ and 49/2$^{-}$ states of DB I, respectively, and shown in Fig. \ref{levsch} (a). These measurements also reveal several new states feeding the lower spin states DB I. The most intense transitions of connecting to the new structure were found to be of 352.7 and 632.4-keV transitions. A cross-over E2 transition of energy 481.0-keV which was tentatively placed between the 51/2$^{(-)}$ and 47/2$^{(-)}$ states in Ref. \cite{rajban7}, has been confirmed as inexistent from the present investigation.  

The investigations have established the MR character of DB I, as indicated in Ref. \cite{rraut}, while DB II, originating out of one of the intermediate states DB I, has been ascribed to collective rotation. The details constitute the findings reported in Ref. \cite{rajban6}. The present paper reports the details of a third band, labelled as DB III in Fig. \ref{levsch} (a), observed in $^{143}$Sm, following the current study. This dipole band DB III, starting at $E_{x}$ = 10081-keV, 49/2$^{-}$ has been extended to an excitation energy of $\approx$ 12 MeV and spin 59/2$^{(-)}$. The intensities of the first observed $\gamma$-ray transition of DB III were extracted from the projection of $E_{\gamma}(90^{\circ})$ vs. $E_{\gamma}(90^{\circ})$ matrix and normalized to that of 239.5-keV transition. The $\gamma$-ray transitions associated with DB III along with their multipolarities, and $R_{DCO}$, $\Delta$ and $\delta$ values are summarized in Table I. The intraband transitions of DB III, 188.9, 352.7, 449.4, 515.0 and 462.1-keV, have been identified to be of mixed $M1$/$E2$ character from the respective $R_{DCO}$ and $\Delta$ values. The cross-over transitions 542.0, 802.0, 965.0 and 977.0-keV, have been confirmed to be of $E2$ nature from the $R_{DCO}$ and $\Delta$ measurements. The cascade DB III was observed to feed the previously reported band DB I principally through the 632.4, and 888.2-keV transitions and to the 51/2$^{-}$ level of DB II through the 457.0-keV transition \cite{rajban4, rajban6, rajban7}. The $M1$/$E2$ nature of these transitions, indicated by the respective $R_{DCO}$ and $\Delta$ values, together with the existing spin-parity assignments of DB I, has facilitated the conclusion on spin-parities of the (feeding) states in DB III.

\begin{figure} [t]
\centering
\setlength{\unitlength}{0.05\textwidth}
\begin{picture}(10,7.2)
\put(-0.2,-0.7){\includegraphics[width=0.55\textwidth, angle = 0]{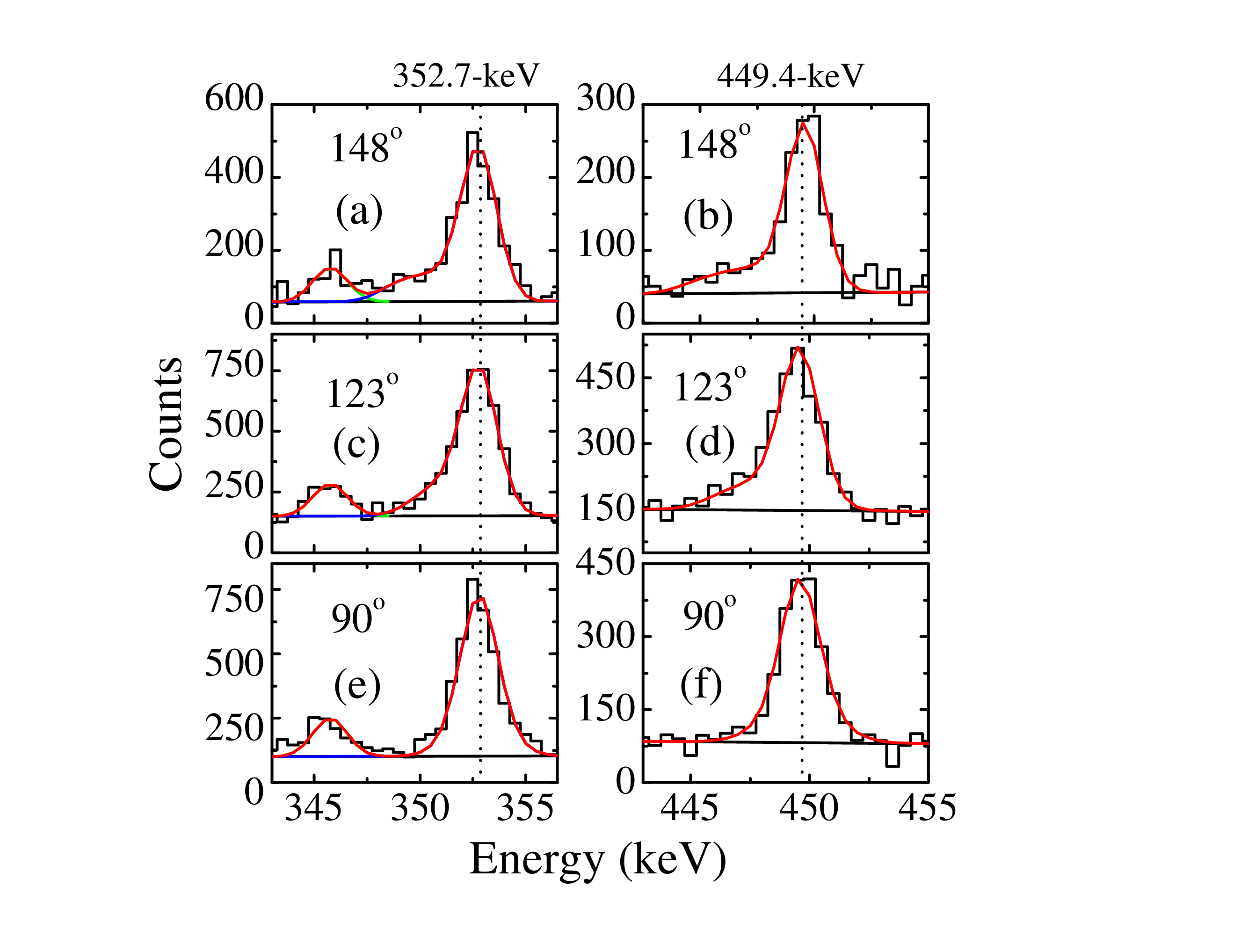}}
\end{picture}
\caption{\label{shape-bot} (Color  online) The panels (a), (c) and (e); and (b), (d), and (f) exhibit the experimental spectra along with the fitted line shapes for the 352.7, and the 449.4-keV $\gamma$ transitions, respectively, of the band DB III in $^{143}$Sm. The top, middle and bottom rows correspond to the shapes in the 148$^{\circ}$, 123$^{\circ}$ and 90$^{\circ}$ detectors, respectively. The obtained line-shape of $\gamma$ transition, contaminant peaks and total line-shapes are represented by the blue, green and red curves, respectively.}
\end{figure}

\begin{figure}[t]
\centering
\setlength{\unitlength}{0.05\textwidth}
\begin{picture}(10,11.0)
\put(-0,-0.4){\includegraphics[width=0.50\textwidth, angle = 0]{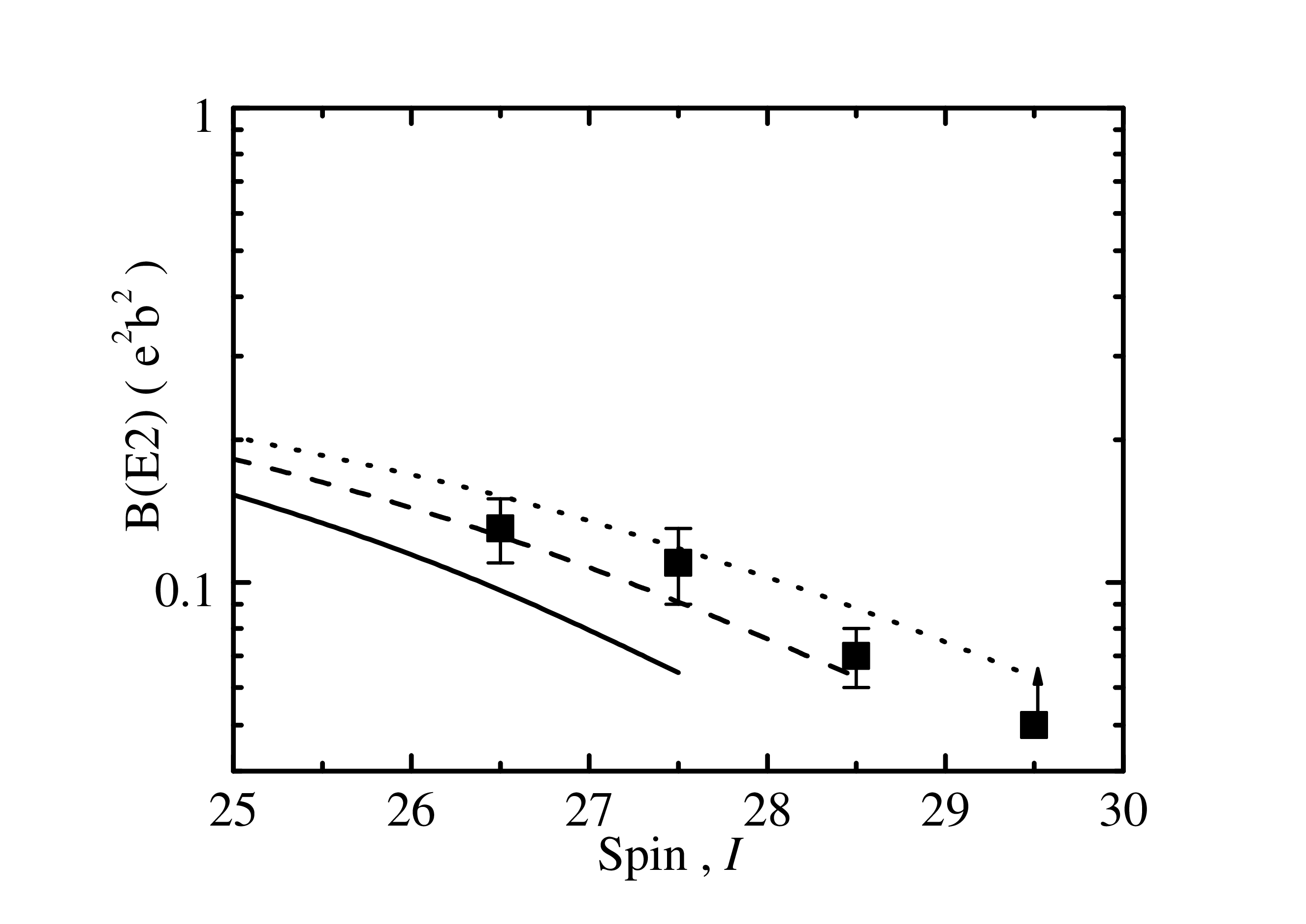}}
\put(-0,+4.85){\includegraphics[width=0.50\textwidth, angle = 0]{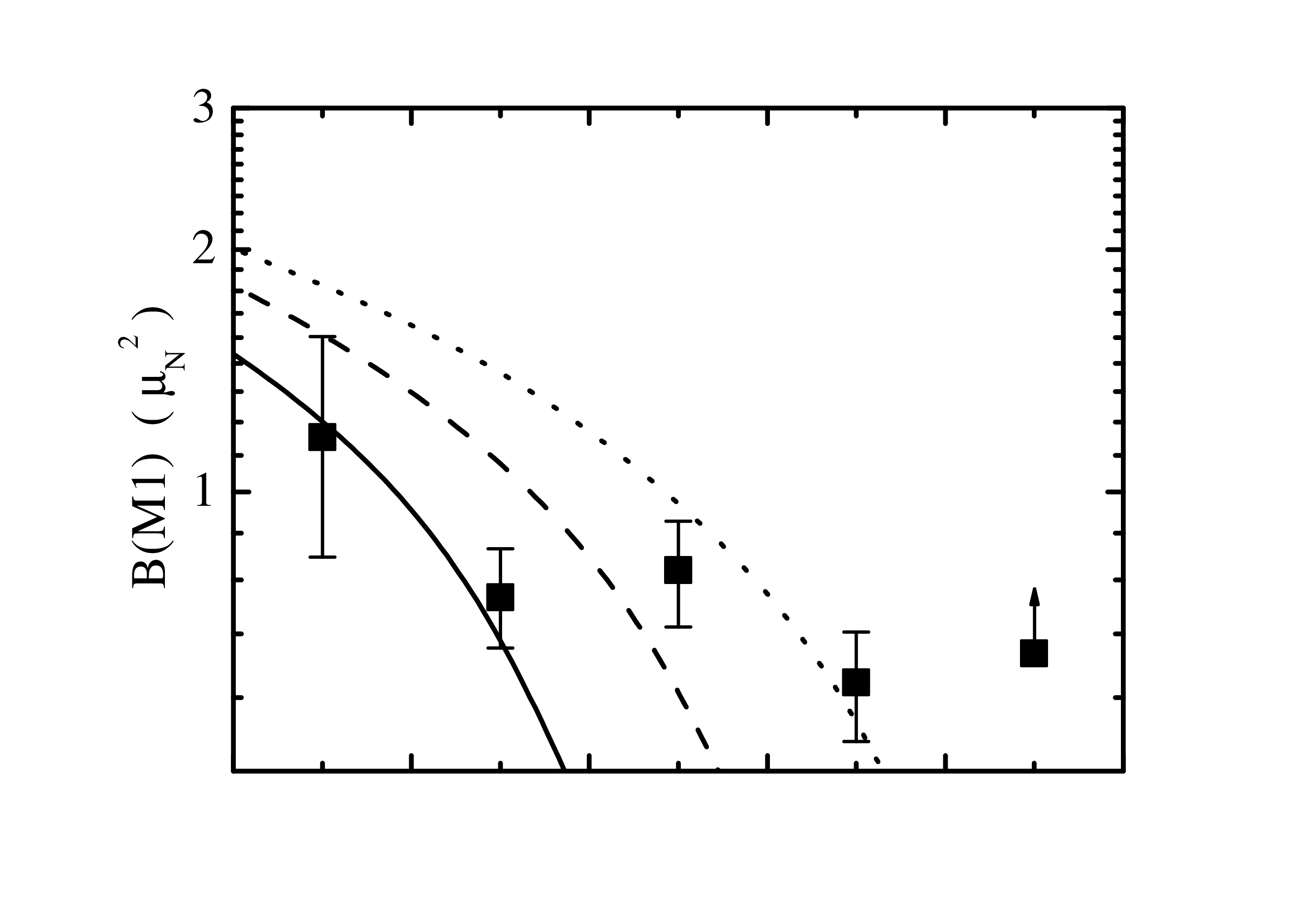}}
\put(+4.5,+7.80){\includegraphics[width=0.21\textwidth, angle = 0]{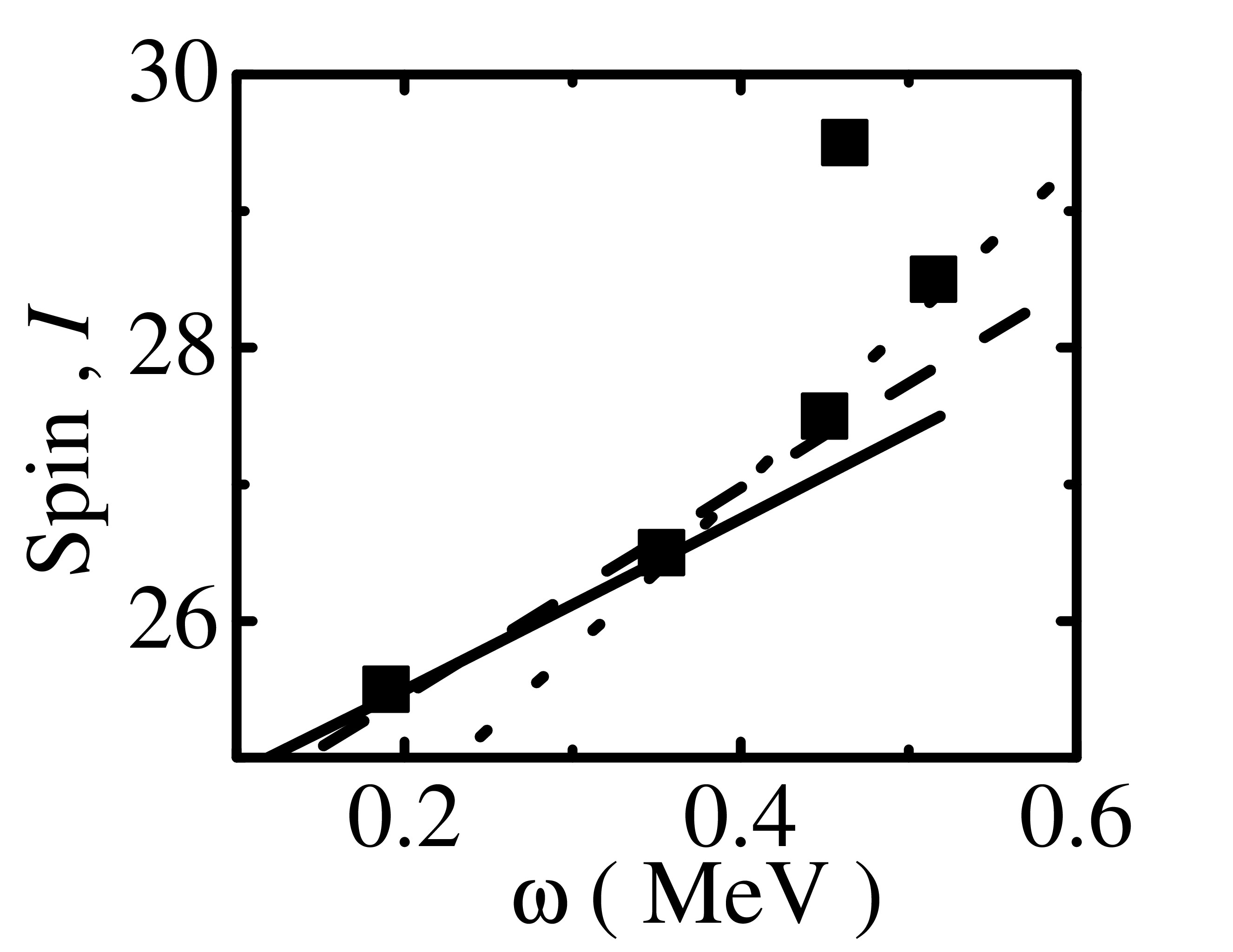}}
\put(+4.2,+2.60){\includegraphics[width=0.22\textwidth, angle = 0]{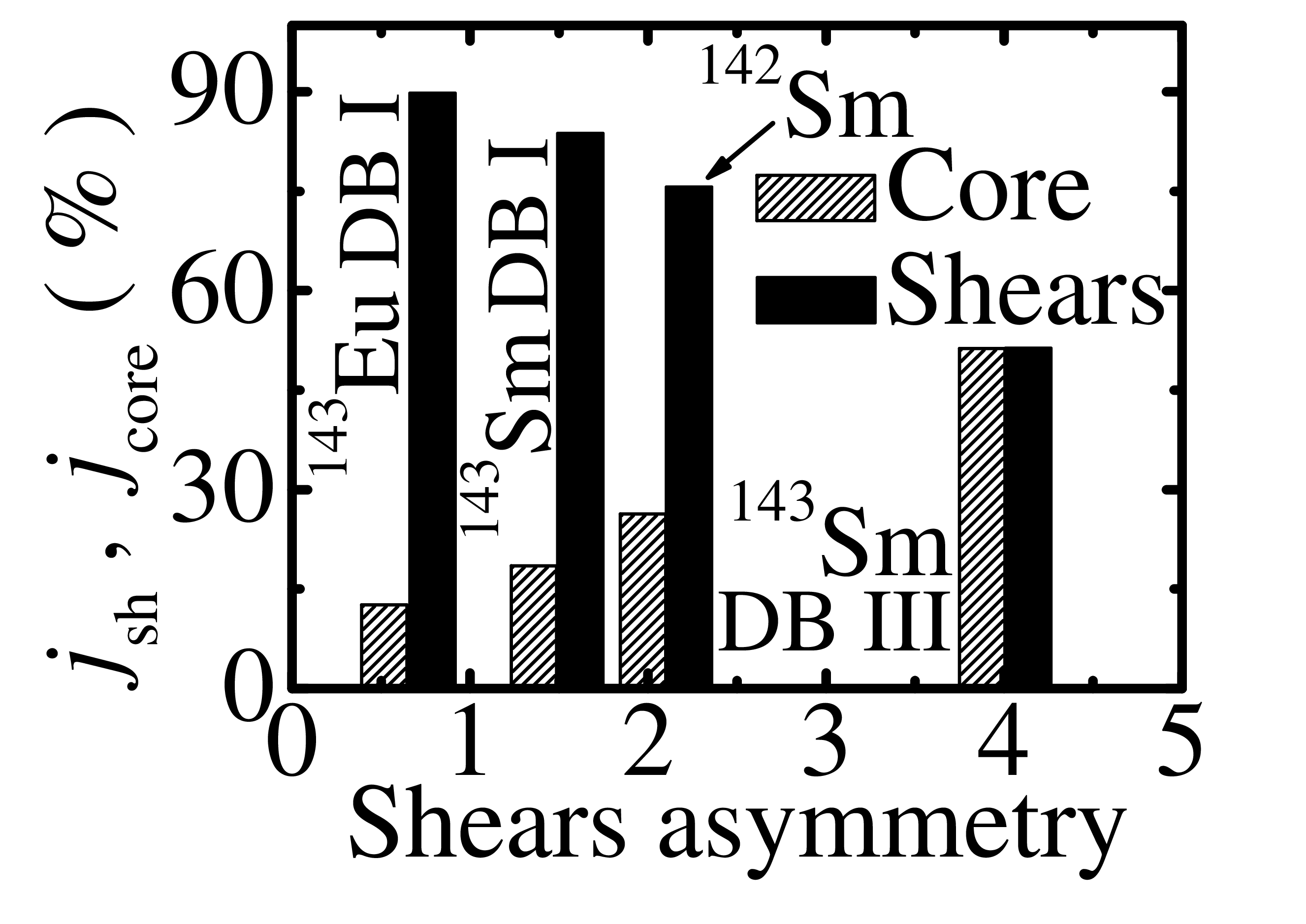}}
\put(2.5,7.0){\textbf{(a)}}
\put(2.5,1.5){\textbf{(b)}}
\end{picture}
\caption{\label{theo-cal} Comparison of the experimental results for the dipole bands DB III (represented by the black filled squares) in $^{143}$Sm with the modified SPAC model calculations for different values of $\chi$ represented by the solid ($\chi$ = 1.0), dash ($\chi$ = 1.2) and dot ($\chi$ = 1.4) black lines. The $B(M1)$ and $B(E2)$ transition strengths against spin ($I$) have been depicted in (a) and (b), respectively. The nature of spin ($I$) against the rotational frequency ($\omega$) is shown in the inset of (a). The inset of (b) shows the variation of the shears and core angular momentum with the asymmetry of the shears blades of the MR bands in the Eu and Sm nuclei.}
\end{figure}

The $\gamma$ -ray transitions of the DB III band were observed to exhibit Doppler shapes in the experimental spectra, that facilitated the determination of level lifetimes and transition probabilities therein. The exercise was based on Doppler shift attenuation method (DSAM) and was carried out using the developments reported in ref. \cite{das16} and the LINESHAPE package \cite{wells, johnson}. The basic procedure is to calculate the Doppler shape of the transitions of interest from stopping simulations, detector geometry and level scheme information, and then least square fit the calculated shape to the experimental spectrum so as to extract the level lifetime. The DSAM analysis has been detailed in Ref. \cite{rajban1, rajban6}. In the present analysis, spectra at 90$^{\circ}$, 123$^{\circ}$ and 148$^{\circ}$ have been fitted simultaneously for extracting the level lifetimes. Representative fits of the calculated Doppler shape to the experimental ones are illustrated in Fig. \ref{shape-bot}. The procedure started with the topmost level, assumed to be 100\% side (top) feed and was continued to the states below, as per the standard methodology. For states feed by both $M1$ + $E2$ transition from the state above and the cross-over $E2$ transition, the intensity weighted average of the lifetimes of the respective feeding states was incorporated in the analysis. For instance, the feeding history of the 55/2$^{-}$ state was represented by intensity weighted average of the lifetimes of the 57/2$^{-}$ and 59/2$^{(-)}$ levels. Likewise, the procedures was adopted for the other levels. The extracted level lifetimes and corresponding sidefeeding lifetimes are depicted in Table I. The $B(M1)$ and $B(E2)$ values derived from the lifetime results are recorded in Table I and  plotted in Fig. \ref{theo-cal}. The values of these transition probabilities and their evolution along the band facilitate an insight into the associated physics, as elaborated hereafter. The uncertainties on lifetime values were derived from the behaviour of $\chi^{2}$ in the vicinity of the minimum. Systematic uncertainty from stopping powers, that is expected to be of $\approx$ 5\% has not been included in the quoted uncertainties (Table I).

The dipole band structure DB III, consisting of the dipole transitions of regular energy spacing, has close similarity to the observed MR bands in the neighbouring weakly deformed nuclei $^{139, 141, 142}$Sm, $^{141,143}$Eu, and $^{142}$Gd \cite{pasern, podsvi, pasern1, rajban1, rajban, rajban3}. The magnitude of the $B(M1)$ transition strength at the 51/2$^{-}$ level of DB III is about three times shorter than the $B(M1)$ value at the 45/2$^{-}$ state of DB I which has been identified as MR band in $^{143}$Sm \cite{rajban4, rajban6}. Also, the $B(M1)$ values are changing slowly from the 1.17$^{+0.39}_{-0.34}$ $\mu_{N}^{2}$ at 51/2$^{-}$ to the 0.58$^{+0.09}_{-0.09}$ $\mu_{N}^{2}$ at 57/2$^{-}$ in contrast to the sharp falling trend for the MR bands observed in this mass region (Table \ref{lifetime} and Fig. \ref{theo-cal} (a)). These characteristics of the $B(M1)$ transition strength are indicative of the deformed rotational character of the band DB III. In contrary, measured $B(E2)$ transitions strengths, in the present case, are comparable to the established MR bands reflects the weakly deformed nature of the band DB III (Table \ref{lifetime} and Fig. \ref{theo-cal} (b)). The small values of $B(M1)$ at the band head and its slowly falling trend along the band, at variance with the established MR sequences in the same nucleus, along with small $B(E2)$ values, indicative of weak deformation, may be perceived as an interplay of the shears mechanism and the collective rotation associated with the DB III sequence in $^{143}$Sm.

The Shears mechanism with the Principal Axis Cranking (SPAC) model \cite{pasern, podsvi, pasern1, rajban1, rajban, rajban3, rajban4, rajban6} has been identified as a powerful tool to explore the intrinsic character, quasiparticle configurations, and contribution of (deformed) core rotation in shears sequences. To achieve an quantitative measures of the core rotational contribution to the total angular momentum vector the SPAC model has been modified and the reduced energy of the state with reduced spin ($\hat{I}$ = $I$/2$j_{1}$) has been expressed as \cite{rajban5},

\begin{center}
$\hat{E}(\hat{I})= E(I)$$\times$$\frac{J(I)}{2j_{1}^{2}} = \hat{I}^{2} + \frac{1 + a^{2}}{4} -  \frac{1}{2}(\sqrt{4\hat{I}^{2} - \sin^{2}\theta_{1}} (a + \cos\theta_{1}) + \sin^{2}\theta_{1}) + \frac{a}{2} \cos\theta_{1} + \frac{\chi}{4} \cos^{2}\theta_{1} - \frac{\chi}{12} $
\end{center}

where, $E(I)$, $J(I)$ and $v_{2}$ are the energy of the state, core moment of inertia and the particle-hole interaction potential, respectively. Here, $\chi$ = $\frac{ J(I)}{j_{1}^2/3v_{2}}$ is a dimensionless quantity, determines the contribution of the core rotation in the shears band and $a$ = $j_{2}$/$j_{1}$ is the asymmetry factor of the angular momentum vectors $j_{1}$ and $j_{2}$, determines the asymmetry of the shears blades. $\theta_{1}$ represents the angle between the angular momentum vectors $\overrightarrow{j_{1}}$ and $\overrightarrow{j_{2}}$ where the direction of $\overrightarrow{j_{2}}$ is set along the rotational axis ($\overrightarrow{R}$).

Minimizing $\hat{E}(\hat{I})$ with respect to the angle $\theta_{1}$, the rotational frequency ($\hat{\omega}$) for the state with reduced spin $\hat{I}$ has been obtained as,

\begin{center}
$\hat{\omega} = 2\hat{I} (1 - \chi + \frac{\sqrt{4\hat{I}^{2} - \sin^{2}\theta_{1}} - a}{\cos\theta_{1}})$ . 
\end{center}

The minimized value of $\theta_{1}$ is then used to determine the $B(M1)$ and $B(E2)$ transition rates \cite{pasern, podsvi, pasern1, rajban1, rajban, rajban3}. The details of this model is outlined in Ref. \cite{rajban5}.

The experimental quasiparticle alignment gain $i_{x}$ $\approx$ 2$\hbar$ of the for the band DB III against the dipole bands DB I and DB II (Fig. \ref{levsch} (c)) is in agreement with the promotion of two protons to the $h_{11/2}$ orbital from the $(g_{7/2}/d_{5/2})$ orbital with respect to the seven quasiparticles ($\pi h_{11/2}^{4}$ $\pi(g_{7/2}/d_{5/2})^{-2}$ ${\otimes}$ $\nu{h}_{11/2}^{-1}$) bands DB I and DB II. Due to the Pauli blocking (for DB I and DB II, four aligned protons are in the $11/2$, $9/2$, $7/2$ and $5/2$ projections of the $h_{11/2}$ orbital) the available projected states for the promoted pair of $h_{11/2}$ protons are $3/2$ and $1/2$ which is corroborated by the experimental alignment. These arguments unambiguously imply the negative parity band DB III has been originated from the configuration $\pi h_{11/2}^{6}$ $\pi(g_{7/2}/d_{5/2})^{2}$ ${\otimes}$ $\nu$$h_{11/2}^{-1}$. Within this configuration the bandhead spin, 49/2$^{-}$, has been reproduced well considering the perpendicular coupling of the hole and particle angular momentum vectors therein. The maximum spin that can be generated from the configuration is 59/2$\hbar$ which is also in good agreement with the observed states of this structure. Furthermore, the experimental values of $B(M1)$, $B(E2)$, and rotational frequency ($\omega$) are well reproduced in theoretical calculations assuming prolate deformation and the unstretched condition of the angular momenta with $j_{1}$ = 4.5$\hbar$, $g_{1}$ = -0.21, $g_{2}$ = +1.21 and a = 4 (Fig. \ref{theo-cal}). The close comparison of the experimental results ($B(M1)$, $B(E2)$, and $\omega$) within the modified SPAC model calculations exhibit that low spin behaviour is well explained by $\chi$ = 1.0 whereas the high spin states are in agreement with $\chi$ = 1.4 (Fig. \ref{theo-cal}) up to the spin 57/2$^{-}$. This calculation under predicts the experimental $B(M1)$, $B(E2)$, and $\omega$ for the 59/2$^{-}$ state, that might be indicative of a new configuration therein. However, as the parameter $\chi$ is a representative of the core contribution to the total angular momentum the calculation exhibits that the core angular momentum smoothly increases from $\approx$ 50\% to the $\approx$ 60\% up to the spin 57/2$^{-}$ of the DB III. The diminishing contribution of the shears angular momentum along the band is correspondingly established. Thus, it may be inferred that the band DB III in $^{143}$Sm may be a maiden example a shears band giving in to the core rotation and exhibiting an interplay of these different mechanisms for generation of angular momentum in the nucleus.

As far as the DB III band is concerned, length of the angular momentum blades of the shears generated within the nine quasiparticle configuration $\pi h_{11/2}^{6}$ $\pi(g_{7/2}/d_{5/2})^{2}$ ${\otimes}$ $\nu{h}_{11/2}^{-1}$ are 24$\hbar$ and 5.5$\hbar$ for particle and hole sectors, respectively, reflects the asymmetric nature of the shears. The inset of Fig. \ref{theo-cal} (b) depicts the variation of shears and core angular momentum with asymmetry of the shears blades of the MR bands in $^{143}$Eu (DB I), $^{143}$Sm (DB I and DB III) and $^{142}$Sm \cite{rajban5}. This is indicative of the increased contribution of the core angular momentum with increasing asymmetry of the angular momentum blades associated with the shears band. Thus, present situation, in case of the dipole band DB III, represents extreme limit of the asymmetric nature of the shears blades above which no shears mechanism would be expected in accordance to the Clark and Macchiavelli \cite{rmcla1}. This is because the angular momentum generated up to the spin 57/2$^{-}$ of the band DB III due to the rotation of core is $\approx$ 50\% of the total angular momentum or more.

In summary, the intrinsic nature of a newly observed dipole band (DB III) in $^{143}$Sm has been investigated. A nine quasiparticle configuration has been assigned to the sequence. Such an observation of this band with, to the best of our knowledge, hitherto highest number of quasiparticles and asymmetricity (among similar structures observed in this and other mass regions) is unique and significant. Also significant is the major contribution of the core angular momentum in the band. It represents an observation of the shears mechanism operational at the boundary wherein the few quasiparticle excitations, such as manifested in the MR phenomenon, evolve into collectivity. This limiting character of the present observation is brought out in experimental results such as the slow falling trend of B(M1), contrary to the more emphatic one typically exhibited by MR sequences, and upholds our contention of this being an example of MR mechanism at the emanation of collectivity. Moreover, such observations across the nuclear chart would facilitate an understanding of this evolution and its dependence on the structural characteristics of the respective mass regions.

\begin{center}
$\textbf{ACKNOWLEDGMENTS}$
\end{center}

We would like to acknowledge the help from all INGA collaborators. We are thankful to the Pelletron staff for giving us steady and uninterrupted $^{24}$Mg beam. S. R. would like to acknowledge the financial assistance from the University Grants Commission - Minor Research Project (No. PSW-249/15-16 (ERO)). G. G acknowledges the support provided by the University Grants Commission - departmental research support (UGC-DRS) program. H. P. is grateful for the support of the Ramanujan Fellowship research grant under SERB-DST (SB/S2/RJN-031/2016).


\begin{thebibliography}{}


\bibitem{ajsim} A. J. Simons \textit{et al.}, Phys. Rev. Lett. {\bf 91}, 162501 (2003).

\bibitem{hubel} H. H$\ddot{u}$bel, Prog. Part. Nucl. Phys. {\bf 54}, 1 (2005).

\bibitem{rmcla1} R. M. Clark and A. O. Macchiavelli, Annu. Rev. Nucl. Part. Sci. {\bf 50} 1 (2000).

\bibitem{amita} Amita, A. K. Jain, and B. Singh, At. Data Nucl. Data Tables {\bf74}, 283 (2000).

\bibitem{frauen1} S. Frauendorf, Rev. Mod. Phy. {\bf 73}, 463 (2001).


\bibitem{108cd} N. S. Kelsall \textit{et al.}, Phys. Rev. C {\bf 61}, 011301(R) (1999).


\bibitem{199pb} A. G$\ddot{o}$rgen \textit{et al.}, Nucl. Phys. {\bf A 683}, 108 (2005).


\bibitem{198bi} H. Pai \textit{et al.}, Phys. Rev. C {\bf 90}, 064314 (2014).




\bibitem{rraut} R. Raut \textit{et al.}, Phys. Rev. C {\bf 73}, 044305 (2006).



\bibitem{rajban4} S. Rajbanshi \textit{et al.}, arXiv:1710.10019 [nucl-ex].

\bibitem{rajban6} S. Rajbanshi \textit{et al.},  Phys. Lett. B {\bf 782}, 143 (2018).

\bibitem{rajban7} S. Rajbanshi \textit{et al.},  Proceedings of the DAE Symp. on Nucl. Phys. 59, 70 (2014).

\bibitem{smurali} S. Muralithar \textit{et al.}, Nucl. Instrum. Methods Phys. Res. {\bf A} 622, 281 (2010).


\bibitem{ingasort} R. K. Bhowmik, Ingasort Manual, private communication. 

\bibitem{radford1} D. C. Radford, Nucl. Instrum. Methods, Phys. Res., Sect {\bf A 361}, 297 (1995).

\bibitem{radford2} D. C. Radford, Nucl. Instrum. Methods, Phys. Res., Sect {\bf A 361}, 306 (1995). 


\bibitem{kramer} A. Kr$\ddot{a}$mer-flecken, T. Morek, R. M. Lieder, W.
 Gast, G. Hebbinghaus, H.M. J$\ddot{a}$ger and W. Urban, Nucl. Instrum.
 Methods. phys. res. {\bf A 275}, 333-339 (1989).

\bibitem{kabadi} M.K. Kabadiyski, K.P. Lieb, D. Rudolph, Nucl. Phys.
 {\bf A 563}, 301-325 (1993).


\bibitem{staro2} K. Starosta \textit{et al.}, Nucl. Instrum. Methods, Phys. Res. A {\bf 423} (1999) 16 - 26.

\bibitem{droste2} Ch. Droste \textit{et al.}, Nucl. Instrum. Methods Phys. Res. A {\bf 378} (1996) 518-525.

\bibitem{deng2} J. K. Deng \textit{et al.}, Nucl. Instrum. Methods Phys. Res. A {\bf 317}, 242 (1992).

\bibitem{jones2} P. M. Jones \textit{et al.}, Nucl. Instrum. Methods Phys. Res. A {\bf 362}, 556 (1995).

\bibitem{tyamaza} T. Yamazaki, Nucl. Data section A 3, 1 (1967).

\bibitem{edmawsu} E.D. Mateosion and A.W. Sunyar, Atomic Data. Nucl. Data Tables 13, 392 (1974).
 

\bibitem{macias} E. S. Macias, W. D. Ruhter, D.C. Camp and R.G. Lanier, Computer Physics Communications {\bf 11}, 75—93 (1976).

\bibitem{dmcull} D.M. Cullen \textit{et al.}, Phys. Rev. C 66, 034308 (2002).

\bibitem{rajban} S. Rajbanshi \textit{et al.}, Phys. Rev. C {\bf 89}, 014315 (2014).

\bibitem{rajban3} S. Rajbanshi \textit{et al.}, Phys. Rev. C {\bf 94}, 044318 (2016).


\bibitem{das16} S. Das \textit{et al.}, Nucl. Instrum. Methods Phys. Res. A {\bf 841}, 17 (2017).

\bibitem{wells} J.C. Wells, N.R. Johnson, LINESHAPE: A Computer Program 
for Doppler Broadened Lineshape Analysis, Report No. ORNL-6689  (1991) 44.

\bibitem{johnson} N.R. Johnson \textit{et al.}, Phys. Rev. C {\bf 55}, 652 (1997).

\bibitem{rajban1} S. Rajbanshi \textit{et al.}, Phys. Rev. C {\bf 90}, 024318 (2014).


\bibitem{pasern} A. A. Pasternak, \textit{et al.}, Eur. Phys. J. {\bf A 37}, 279 - 286 (2008).

\bibitem{podsvi} E. O. Podsvirova \textit{et al.}, Eur. Phys. J.
 {\bf A 21}, 1 - 6 (2004).

\bibitem{pasern1} A. A. Pasternak \textit{et al.}, Eur. Phys. J.
 {\bf A 23}, 191 - 196 (2005).

\bibitem{rajban5} S. Rajbanshi, arXiv:1712.00353v1 [nucl-th].


\end{thebibliography}
\end{document}